\documentclass[12pt]{article}
\usepackage{psfig}

\begin{document}

\author{G. Herrera$^{1\footnote{gherrera@fis.cinvestav.mx}}$, 
J. Magnin$^{2\footnote{jmagnin@cbpf.br}}$
L.M. Monta\~no$^{1\footnote{lmontano@fis.cinvestav.mx}}$\\
\small{$^1$ Depto. de F\'{\i}sica del Centro de Investigaci\'on y de Estudios Avanzados del IPN}\\
\small{Apartado postal 14-740, M\'exico D.F. 07000, M\'exico}\\
\small{$^2$ Centro Brasileiro de Pesquisas F\'{\i}sicas}\\
\small{Rua Dr. Xavier Sigaud 150 - CEP 22290-180, Rio de Janeiro, Brazil}}

\title{Longitudinal $\bar \Lambda_0$ polarization in heavy ion collisions as 
a probe for QGP formation.}

\date{\today}
\maketitle

\begin{abstract}
We present an analysis of the longitudinal $\bar \Lambda_0$ polarization 
in ultra-relativistic heavy-ion collisions. The polarization of $\bar \Lambda_0$'s coming 
from the decay chain $\bar{\Xi}\rightarrow \bar{\Lambda}_0+\pi$ exhibits a very well 
differentiated behavior depending on the production region of the primordial 
$\bar \Xi$'s. This effect reflects the different values of the 
$N_{\bar \Xi}/N_{\bar \Lambda_0}$ ratio in the QGP region, where nucleon-nucleon 
interactions take place in a hot and dense environment, 
and the peripherical region, in which ordinary nucleon-nucleon interactions occur.
An increase in the longitudinal $\bar \Lambda_0$ polarization signals a strangeness 
enhancement which is thought as a property of the QGP phase.
\end{abstract}

\section{Introduction}

In nuclear collisions at relativistic and ultra-relativistic energies, it is 
expected a phase transition from ordinary nuclear matter to a Quark Gluon Plasma 
(QGP), which should be observed when sufficiently high baryonic densities and/or 
temperatures be achieved  in the collision. In order to identify this phase transition, 
a number of experimental observables, namely $J/\Psi$ suppression, strangeness enhancement, 
fluctuations in particle ratios, flow patterns, etc.  have been proposed~\cite{observables}. 

In particular, it has been argued that the strangeness enhancement in hot and dense regions 
of nuclear matter would lead to an abundant formation of multi-strange baryons and antibaryons, 
providing them a key information about the QGF formation~\cite{rafelski}. Indeed, detailed 
calculations~\cite{calculos} predict that the abundance of $\bar{\Xi}(\bar s\bar s \bar q)$ 
should be enriched to about half the abundance of antihyperons $\bar Y(\bar s \bar q \bar q)$ 
as compared to the $\bar {\Xi}/\bar {Y}$ ratio seen in nucleon-nucleon interactions. 
Considering that at $\sqrt{s} = 63$ GeV, $\bar {\Xi}/\bar {\Lambda_0} = 
0.06\pm0.02$~\cite{isr} in the central rapidity region, then in the presence of QGP, 
the $\bar {\Xi}/\bar {\Lambda_0}$ would be ten times greater. However, although the 
$\bar {\Xi}/\bar {\Lambda_0}$ ratio is a quantity which is 
difficult to establish experimentally, the longitudinal $\bar \Lambda_0$ polarization 
is not and can be used to get a measurement of the above mentioned ratio~\cite{rafelski}.

At this point it is useful to remember that $\bar \Xi$ decays into $\bar \Lambda_0 + \pi$ with 
a branching fraction of about $99~\%$ and that the weak decay polarizes 
the $\bar \Lambda_0$ spin longitudinally. This means that all the longitudinally 
polarized $\bar \Lambda_0$ are associated with the primordial abundance of $\bar \Xi$. 
Note also that $\bar\Omega^+$ have a little influence on the particle abundances and, in 
particular, over their polarizations~\cite{rafelski}. In fact, the STAR 
Collaboration~\cite{star1} measured 
$N_{\Omega}/N_{\bar\Xi^+}\sim 0.16$, where $N_\Omega$ refers to the total number of 
$\Omega^-+\bar\Omega^+$. Furthermore, the polarization of $\bar\Lambda_0$'s coming from 
$\bar\Omega^+ \rightarrow \Lambda_0 + K^+$ (BR $\sim$ 68\%) is a factor of 5 lower than 
polarization of $\bar\Lambda_0$'s coming from $\bar\Xi \rightarrow \bar\Lambda_0+\pi$ while 
the polarization of $\bar\Lambda_0$'s coming from the decay 
$\bar\Omega^+ \rightarrow \bar\Xi + \pi$ (BR $\sim$ 23\% and BR $\sim$ 9\% for the 
$\bar\Xi^0$ and the $\bar\Xi^+$ decay modes respectively) and the subsequent decay of the 
$\bar\Xi$ into $\bar\Lambda_0 + \pi$ is still lower by a factor of $\sim$ 20. This situation 
is expected to be mantained at the energy where QGP formation takes place (see also discussions 
in Refs.~\cite{rafelski,calculos}).

The $\bar \Lambda_0$ polarization can be defined in terms of the so called up-down 
asymmetry of the $\bar \Lambda_0$ decay with reference to the plane normal to the 
$\bar \Lambda_0$ momentum. Thus~\cite{rafelski}
\begin{equation}
\frac{N_u-N_d}{N_u+N_d} = \frac{1}{2}\alpha_{\bar \Lambda} p_{\bar \Lambda}\; ,
\label{eq1}
\end{equation}
where $p_{\bar \Lambda} = \alpha_{\bar \Xi} = -\alpha_\Xi$ is the $\bar \Lambda_0$ polarization 
and $\alpha_{\bar \Lambda}$ is the $\bar \Lambda$ decay parameter. From data tables 
$\alpha_{\bar \Lambda}=-\alpha_\Lambda=0.642\pm0.013$ and $\alpha_{\bar \Xi}=0.413\pm0.022~
(0.455\pm0.015)$ for $\bar \Xi^0~(\bar \Xi^-)$. Hence the total up-down asymmetry for all 
the neutral $\bar \Lambda_0$ events is
\begin{equation}
\frac{N_u-N_d}{N_u+N_d} = \gamma
\frac{1}{2}\alpha_{\bar \Lambda} \alpha_{\bar \Xi}\; ,
\label{eq2}
\end{equation}
where $\gamma = N_{\bar \Xi}/N_{\bar \Lambda_0}$.

Eq.~(\ref{eq2}) shows that in ordinary nucleon-nucleon interactions at the ISR energies, the 
up-down asymmetry should be of the order of $-0.008$ while in hot and dense nuclear matter 
it should amount to something about $-0.07$, which is a effect of one order of magnitude and 
that could be measured in experiments.

Of course, all of the above argumentation can be applied to $\Lambda_0$ and $\Xi$'s, but in 
this case, many of the $\Lambda_0$'s should be mere fragments of nucleons going into 
$\Lambda_0K$, giving a less clear signal.

\begin{figure}[t]
\centerline{\psfig{figure=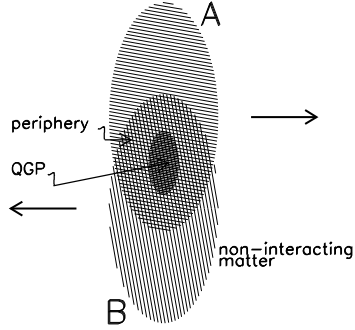,height=3.0in}}
\caption{Schematic representation of the reaction $A+B$ and the regions in which $\Xi$'s  
are produced. In the hot and dense (QGP) region, $\gamma_{QGP} \sim 0.5$ while in the periphery, 
in which ordinary nucleon-nucleon interactions take place, $\gamma_{periph.} << \gamma_{QGP} $.}
\label{fig1}
\end{figure}

In the interaction region of a heavy-ion collision it is expected to have a core of hot and 
dense nuclear matter - possibly QGP - sorrounded by a region in which ordinary nucleon-nucleon 
interactions take place (see Fig.~\ref{fig1}). These regions, as it is shown in the following 
sections, can be mapped by measuring the $\bar{\Lambda}_0$ polarization as a function of the 
impact parameter, $b$, and transverse momentum, $p_T$. What is expected is a plot displaying a 
polarization approximately constant and of the order of $-0.07$ in the hot and dense region 
sorrounded by a region in which the polarization is, again constant, and of the order of $-0.008$, 
corresponding to the periphery, where ordinary nucleon-nucleon interactions occur.

The paper is organized as follows. In Section 2 we study the behavior of the 
$N_{\bar \Xi}/N_{\bar \Lambda_0}$ ratio in both the low and high nuclear density regions. In 
Section 3 we calculate the $\bar{\Lambda}_0$ polarization as a function of the impact parameter 
and transverse momentum, and Section 4 is devoted to conclusions and final remarks.

\section{The $N_{\bar{\Xi}}/N_{\bar{\Lambda}}$ ratio in heavy ion collisions}

In the interaction region of the collision of nucleus $A$ and $B$, when QGP 
coexists with the ordinary nuclear matter, the ratio of the number of $\bar \Xi$ to the 
number of $\bar \Lambda_0$ baryons is given by
\begin{eqnarray}
\gamma_{A+B} &=& \frac{N_{\bar \Xi}^{periph.} + N_{\bar\Xi}^{QGP}}
{N_{\bar \Lambda_0}^{periph.} + N_{\bar \Lambda_0}^{QGP}}\nonumber \\ 
&=& \frac{N_{\bar \Lambda_0}^{periph.}\gamma_{periph.} + N_{\bar \Lambda_0}^{QGP}\gamma_{QGP}}
{N_{\bar \Lambda_0}^{periph.} + N_{\bar \Lambda_0}^{QGP}}\; ,
\label{eq4}
\end{eqnarray}
where the quantities labeled $^{periph}$ and $^{QGP}$ refer respectively to the 
number of $\bar{\Lambda}_0$ and $\bar{\Xi}$ in each region. As long as the dependence 
on the impact parameter, $b$, and transverse momentum, $p_T$, of $N_{\bar \Lambda_0}$ is 
different in the QGP than in the peripherical region, $\gamma_{A+B}$ must be also dependent on 
$b$ and $p_T$. In fact, in the same way as the longitudinal polarization of $\bar{\Lambda}_0$'s 
does, it should provide a map of the interacting region in which the peripherical and QGP 
zones are displayed. Note, however, that the only dependence on $b$ and $p_T$ in $\gamma_{A+B}$ 
arises from the dependence on $b$ and $p_T$ in $N_{\bar \Lambda_0}^{periph.}$ and 
$N_{\bar \Lambda_0}^{QGP}$ since $\gamma_{periph.}$ and $\gamma_{QGP}$ are expected to be 
approximately constant~\cite{isr}.

\subsection{$N_{\bar{\Lambda}_0}(b,p_T)$ in the peripherical region}

The $N_{\bar \Lambda_0}^{periph.}$ can be estimated along the lines in Ref.~\cite{ayala}. 
As stated in~\cite{ayala}, and remembering that in ordinary nucleon-nucleon interactions 
the typical behavior of the production cross section as a function of $p_T$ is an exponential
in $p_T^2$, the number of produced $\bar \Lambda_0$'s  as a function of 
the impact parameter and $p_T$ in the collision of nucleus A and B can be written as
\begin{equation}
\frac{d^4 N^{periph.}_{\bar \Lambda_0}}{d^2b~dp_T^2} = \frac{1}{2}
Ce^{-ap_T^2}T_{AB}(b)\; ,
\label{eq5}
\end{equation}
where the $1/2$ is because the number of $\bar \Lambda_0$'s is approximately half the 
number of neutral hyperons, $C$ is a constant which normalizes the 
integral over the transverse momentum to unity, and $T_{AB}$ is
\begin{equation}
T_{AB}(z,{\bf b})=\int{d^2s~T_A(z,{\bf b})T_B(z,{\bf{s-b}})}\; ,
\label{eq6}
\end{equation}
with $T_A$ and $T_B$ given by
\begin{equation}
T_A(z,{\bf s})=\int_{-z/2}^{z/2}{dz'~\rho_A(z',{\bf s})}\; ,
\label{eq7}
\end{equation}
and where the limits of integration over $z$ have to be extended to 
$\left[-\infty,+\infty \right]$. For $\rho_A$, which is the nucleon density per unit 
area in the transverse plane with respect to the collision axis, we use the standard 
Woods-Saxon density profile
\begin{equation}
\rho_A({\bf r})=\frac{\rho_0}{1+e^{(r-R_A)/d}}\; ,
\label{eq8}
\end{equation}
with $R_A=1.1A^{1/3}$ fm, $d = 0.53$ fm~\cite{povh} and $\rho_0$ fixed by normalization
\begin{equation}
\int{d{\bf r}~\rho_A({\bf r})}=A\; ,
\label{eq9}
\end{equation}
giving $\rho_0=0.17$ fm$^{-3}$ in the case of $^{197}$Au, which we shall consider as an 
example in the following.

We are assuming that each peripheral collision produces final state particles in the same 
way than in free nucleon reactions. However, in order to exclude the zone where the density 
of participants $n_p$ is above the critical density $n_c$ to produce QGP, we rewrite 
eq.~(\ref{eq5}) as
\begin{equation}
\frac{d^4 N^{periph.}_{\bar \Lambda_0}}{d^2b~dp_T^2} =\frac{1}{2} 
Ce^{-ap_T^2}\int{d^2s~T_A(z,{\bf b})T_B(z,{\bf{s-b}})
\Theta\left[n_c-n_p({\bf s},{\bf b})\right]}\; ,
\label{eq10}
\end{equation}
where $n_p({\bf s},{\bf b})$ is the density of participants at the point ${\bf s}$ and 
$\Theta$ is the step function. The density of participants per unit transverse area in the 
collision of nucleus $A$ with nucleus $B$ at an impact parameter ${\bf b}$ has a profile 
given by~\cite{blaizot}
\begin{equation}
n_p(s,{\bf b})=
T_A({\bf s}) \left[ 1-e^{-\sigma_{NN} T_B({\bf s- b})} \right] +
T_B({\bf s- b}) \left[ 1-e^{-\sigma_{NN} T_A({\bf s})} \right],
\label{eq11}
\end{equation}
where $\sigma_{NN}$ is the nucleon-nucleon inelastic cross section which we take as 
$\sigma_{NN}= 32$ mb. The total number of participants $N_p$ at an impact parameter $b$ 
is
\begin{equation}
N_p(b) = \int{d^2s~n_p({\bf s},{\bf b})}\; .
\label{eq12}
\end{equation}

From eq.~(\ref{eq10}), the number of $\bar\Lambda_0$'s coming from the decay chain 
$\bar \Xi\rightarrow \bar \Lambda_0 + \pi$ is then given by
\begin{equation}
\frac{d^4 N^{periph.}_{\bar\Xi\rightarrow \bar\Lambda_0+\pi}}{d^2b~dp_T^2} = 
\gamma_{periph.}\frac{d^4 N^{periph.}_{\bar \Lambda_0}}{d^2b~dp_T^2}\; .
\label{eq12b}
\end{equation}

Following Ref.~\cite{blaizot}, we choose $n_c=3.3$ fm$^{-2}$ for the critical density. 
This number results from the observation of a substantial reduction of the $J/\Psi$ yield 
in Pb - Pb collisions at the SPS. For the parameter $a$ in
eqs.~(\ref{eq5}) and (\ref{eq10}) there is no published data at the relevant energies, 
which for LHC should be of about $5$ TeV per nucleon. Then we use the value measured by 
the Hera-B Collaboration~\cite{hera} in proton-nucleus interactions at 920 GeV, 
$a=2.2\pm0.3$ GeV$^{-2}$.

\subsection{$N_{\bar{\Lambda}}(b,p_T)$ in the QGP region}

In QGP, the average number of produced antistrange quarks scales with the number of participants 
$N^{QGP}_p$ in the collision roughly as~\cite{letessier}
\begin{equation}
\frac{\left<\bar s\right>}{N^{QGP}_p}=cN^{QGP}_p\; .
\label{eq13}
\end{equation}
Assuming that a $\bar s$ quark will produce $\bar \Xi$, $\bar \Lambda_0$ or
$\bar \Sigma_0$, and taken the number of $\bar \Lambda_0$ approximately equal to the 
number of produced $\bar \Sigma_0$, we have
\begin{eqnarray}
\left< \bar s \right> &=& 2 N^{QGP}_{\bar \Xi} + 2 N^{QGP}_{\bar \Lambda_0}\nonumber \\
&=& \left[2\gamma_{QGP} + 2 \right]N^{QGP}_{\bar \Lambda_0}\; .
\label{eq14}
\end{eqnarray}
Combining eqs.~(\ref{eq13}) and (\ref{eq14}) we obtain
\begin{equation}
N^{QGP}_{\bar \Lambda_0} =
\frac{c}{2\gamma_{QGP}+2}\left[N^{QGP}_p\right]^2\; .
\label{eq15}
\end{equation}
$N^{QGP}_p$ as a function of the impact parameter is given, using 
eq.~(\ref{eq12}), as
\begin{equation}
N^{QGP}_p(b)=\int{d^2s~n_p({\bf s},{\bf b})\theta
\left[n_p({\bf s},{\bf b})-n_c\right]}\;.
\label{eq16}
\end{equation}

The eq.~(\ref{eq15}) represents the behavior of the number of $\bar \Lambda_0$ as a function of 
the impact parameter, then  we use
\begin{equation}
\frac{d^4 N^{QGP}_{\bar \Lambda_0}}{d^2b~dp_T^2}=
\frac{c}{2\gamma_{QGP}+2}\left[N^{QGP}_p(b)\right]^2C~'e^{-a'p_T^2}\; ,
\label{eq17}
\end{equation}
assuming an exponential dependence in $p_T^2$ for $\bar \Lambda_0$ production~\cite{star}. As 
for the peripherical $\bar\Lambda_0$'s, eq.~(\ref{eq17}) times $\gamma_{QGP}$ gives the 
number of $\bar \Lambda_0$'s as a function of $b$ and $p_T$  coming from the 
decay chain $\bar \Xi \rightarrow \bar \Lambda_0+\pi$ in the QGP phase. 
We use $c=0.005$~\cite{ayala} and read $a'=0.67$ from Ref.~\cite{star}.

\begin{figure}[t]
\centerline{\psfig{figure=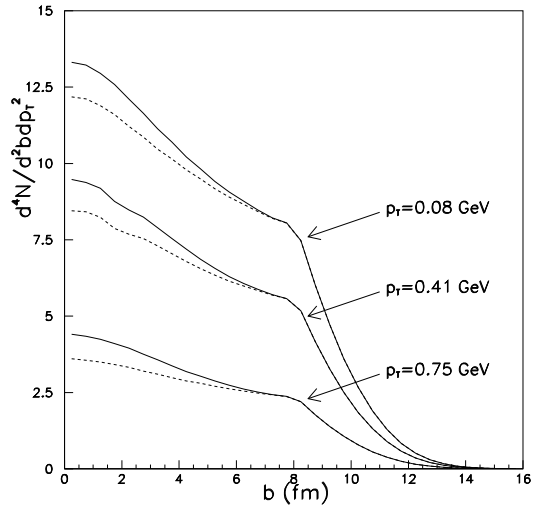,height=3.0in}}
\caption{Number of $\bar\Lambda_0$'s as a function of the impact parameter for 
fixed values of $p_T$. Full line shows the total (peripherical + QGP) number. Dashed line 
is the peripherical contribution.}
\label{fig2}
\end{figure}

\begin{figure}[t]
\centerline{\psfig{figure=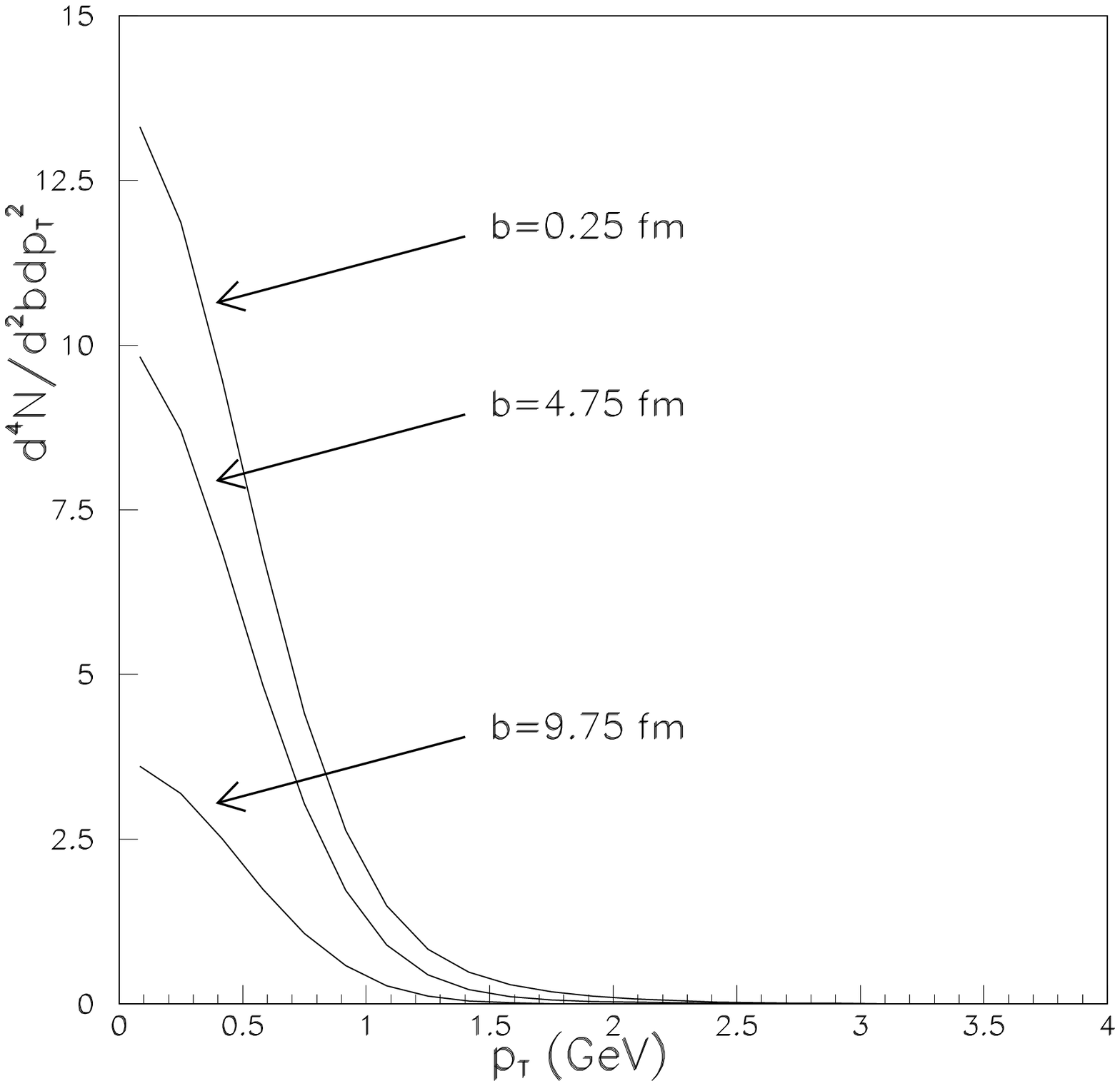,height=3.0in}}
\caption{Number of $\bar\Lambda_0$'s as a function of $p_T$ at fixed values of $b$.}
\label{fig3}
\end{figure}

In Figs. \ref{fig2} and \ref{fig3} we show the behavior of the total (peripherical + QGP) 
number of $\bar \Lambda_0$'s as a function of the impact parameter for different values of $p_T$ 
and as a function of $p_T$ for several values of $b$ respectively.

\section{$\bar{\Lambda}$ polarization in heavy ion interactions}

\begin{figure}[t]
\centerline{\psfig{figure=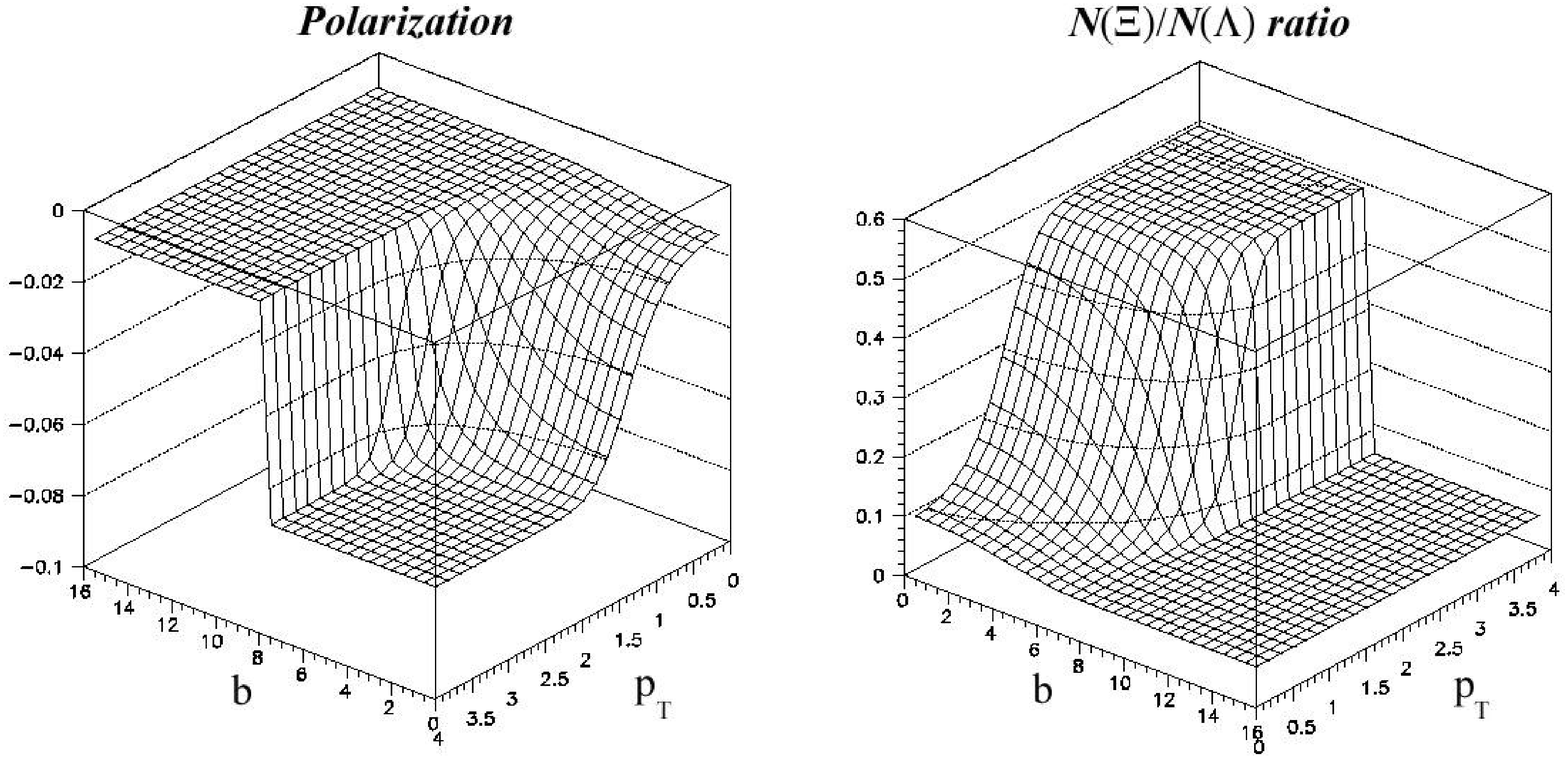,height=2.7in}}
\caption{Left: $\bar{\Lambda}_0$ polarization as a function of $b$ and $p_T$ as defined 
in eq.~\ref{eq2}. We used $\gamma^{perip.}=0.06$ and $\gamma^{QGP}=0.5$ to obtain the plot. 
Right: $\gamma_{eff}=N(\bar{\Xi})/N(\bar{\Lambda_0})$ as a function of $b$ and $p_T$ as 
defined in eq.~\ref{eq4}. Note that this plot is rotated by $180^\circ$ with respect to 
the plot of the polarization for a better visualization. The maximum of $\gamma_{eff}$ 
correspond to the minimum of the $\bar{\Lambda}_0$ polarization.}
\label{fig4}
\end{figure}

The up-down asymmetry is then given by eq.~(\ref{eq2}) with $\gamma_{eff}$ as given in 
eq.~(\ref{eq4}) and shown in the left side of Fig.~\ref{fig4} as a function of $b$ and $p_T$. 
The right side of Fig.~\ref{fig4} shows the behavior of $\gamma_{eff}=
N(\bar{\Xi})/N(\bar{\Lambda_0})$ defined in eq.~\ref{eq4} as a function of $b$ and $p_T$. 
As can be seen, the $\bar \Lambda_0$ polarization exhibits a dramatical change as 
the impact parameter becomes bigger than a critical 
value, starting from which the QGP region suddenly vanishes. From this value on the polarization 
reaches the characteristical value seen in ordinary nucleon-nucleon interactions. Conversely, 
at high $p_T$ and low $b$, the polarization shows the behavior expected in the hot and high 
density region where QGP takes place. This reflects the characteristic increase in the 
$N_{\bar \Xi}/N_{\bar \Lambda_0}$ expected when QGP be formed. Another interesting feature of the 
longitudinal $\bar \Lambda_0$ polarization is the dependence on $p_T$ due 
to the different $p_T$ dependence of $\bar \Lambda_0$ production in the peripherical 
and QGP region.

\begin{figure}[t]
\centerline{\psfig{figure=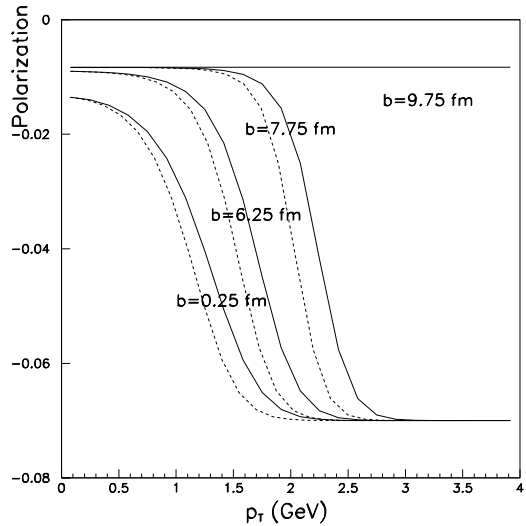,height=3.0in}}
\caption{$\bar \Lambda_0$ polarization as a function of $p_T$ at several values of the 
impact parameter. Full line shows the polarization before the momentum shift due to multiple 
scattering in the nuclear medium, dashed line shows the polarization after the momentum shift.}
\label{fig7}
\end{figure}

However, in relativistic heavy-ion collisions, there are several effects which can modify 
the $\bar \Lambda_0$ polarization and have to be taken properly into account. These 
are i) $\bar \Lambda_0$'s produced by secondary pion-nucleon scattering, ii) secondary 
scattering of $\bar \Lambda_0$'s with nucleons in the interaction region and iii) spin-flip in 
secondary interactions of the $\bar \Lambda_0$'s. $\bar \Lambda_0$'s coming from secondary 
pion-nucleon interactions will have a characteristic low momentum signature and can be 
eliminated from analysis by setting kinematical constrains in the reconstruction of data, 
therefore excluding them from the polarization analysis. Spin-flip effects in polarization, 
which are associated to spin-spin interactions among the $\bar\Lambda_0$ and the surrounding 
particles, are characterized by a polarization transfer coefficient which express the final 
polarization in terms of the initial one as
\begin{equation}
P'= D P\; .
\label{eq18}
\end{equation}
However, due to the lack of experimental information at the range of energies of interest, 
we will omit this effect from our analysis (for a more detailed analysis and Refs., 
see Ref.~\cite{ayala}). Secondary scattering of $\bar \Lambda_0$'s with nucleons in the 
surrounding nuclear environment will produce a momentum shift that can be characterized 
in terms of a sequential model. The final effect will be that, a $\bar\Lambda_0$ 
produced with an initial $(p_L,p_T)$, after multiple-scattering, in the high energy limit, 
will have an average momentum~\cite{ayala}
\begin{eqnarray}
\left<p_L(b) \right> &=& p_L~e^{-I{\bar N}(b)}\nonumber \\
\left<p_T(b)\right> &=& p_T~e^{-I{\bar N}(b)}\cos{\left[\Gamma\sqrt{{\bar N}(b)}\right]}\; ,
\label{eq19}
\end{eqnarray}
where $I=0.2$ is the inelasticity coefficient, $\Gamma=0.01$ is the average dispersion angle
in each collision and ${\bar N}(b)$ is the average number of $\bar \Lambda_0$ collisions in 
the nuclear medium,
\begin{equation}
\bar{N}(b) = \sigma^{tot}_{\bar \Lambda_0 N}~T_A(b/2)\; ,
\label{eq20}
\end{equation}
where $\sigma^{tot}_{\bar \Lambda_0 N}=1.4$ mb~\cite{hera}, taken from proton-nucleus 
interactions at 920 GeV. This effect is shown in Figure ~\ref{fig7} where the polarization 
as a function of $p_T$ is displayed for several values of $b$ before and after the momentum 
shift due to multiple scattering. Note also that, for low $p_T$, the polarization takes the 
value corresponding to that expected in ordinary nucleon-nucleon interactions, a behavior which 
is more evident as $b$ grows. Conversely, at high $p_T$ and low $b$, the polarization reaches 
the value expected in the QGP phase.

\section{Final remarks and conclusions}

In conclusion, we have shown that the longitudinal polarization of $\bar \Lambda_0$ in 
heavy-ion relativistic interactions presents a dramatical change with 
$N_{\bar \Xi}/N_{\bar \Lambda_0}$ in the transition from the expected QGP region to the 
peripherical one. This effect, which can be easily measured in the laboratory can 
help to unveil one of the signals of the transition to the QGP phase, namely the
strangeness enhancement and the consequent increase in the multistrange baryon formation. 
Notice that the longitudinal polarization of the $\bar \Lambda_0$ in the low $b$, high 
$p_T$ region gives direct access to the measurement of the $\gamma_{QGP}$ ratio. It is 
simply the quotient of the polarization by $\alpha_{\bar \Lambda}\alpha_{\bar \Xi}/2$. 
Conversely, in the high $b$, high $p_T$ region, the $\bar \Lambda_0$ longitudinal polarization 
gives the $\gamma_{periph.}$ ratio, which is not easily measured in experiments.
It is interesting to note that, the momentum shift due to multiple-scattering of 
$\bar \Lambda_0$'s in the nuclear medium produces a small increase in the polarization 
as a function of $p_T$ in the low momentum region.

Another possible observable which would exhibit strong changes in the presence of the QGP phase 
is the transverse polarization of the $\Lambda_0$'s~\cite{ayala,griego}. However, in this case 
it is expected a decrease of its value with respect to what happens in p-p interactions since
transverse polarization is related to the production mechanisms of the $\Lambda_0$.

\section*{Acnowledgments}

This work was supported by a binational research agreement CNPq/CONA\-CYT under grant numbers 
690176/02-3 and J200-652. J.M is grateful for the warm hospitality at the Physics Departament 
of CINVESTAV, where part of this work was done. L.M.M. is grateful for the kind hospitality 
at CBPF during the completion of this work




\end{document}